\documentclass[12pt,preprint]{aastex}
\slugcomment{}

\begin{document}

\title{Constraints on cosmological models from lens redshift data}

\author{Shuo Cao and Zong-Hong Zhu}

\affil{Department of Astronomy, Beijing Normal University, Beijing
100875, China; zhuzh@bnu.edu.cn}

\begin{abstract}
Strong lensing has developed into an important astrophysical tool
for probing both cosmology and galaxies (their structures,
formations, and evolutions). Now several hundreds of strong lens
systems produced by massive galaxies have been discovered, which may
form well-defined samples useful for statistical analyses. To
collect a relatively complete lens redshift data from various large
systematic surveys of gravitationally lensed quasars and check the
possibility to use it as a future complementarity to other
cosmological probes. We use the distribution of
gravitationally-lensed image separations observed in the Cosmic Lens
All-Sky Survey (CLASS), the PMN-NVSS Extragalactic Lens Survey
(PANELS), the Sloan Digital Sky Survey (SDSS) and other surveys,
considering a singular isothermal ellipsoid (SIE) model for galactic
potentials as well as improved new measurements of the velocity
dispersion function of galaxies based on the SDSS DR5 data and
recent semi-analytical modeling of galaxy formation, to constrain
two dark energy models ($\Lambda$CDM and constant $w$) under a flat
universe assumption. We find that the current lens redshift data
give a relatively weak constraint on the model parameters. However,
by combing the redshift data with the baryonic acoustic oscillation
peak and the comic macrowave background data, we obtain more
stringent results, which show that the flat $\Lambda$ CDM model is
still included at 1$\sigma$.
\end{abstract}


\section{Introduction}
\label{sec:intro}

The discovery of strong gravitational lensing in Q0957+561 (Walsh et
al.1979) opened up the vast possibility to use strong lens systems
in the study of cosmology and astrophysics. Up to now, strong
lensing has developed into an important astrophysical tool for
probing both cosmology \citep{Fuk92,Koc93,Zhu00a,Zhu00b,Cha02,Cha03,
Cha04a, Mit05, Yor05, Zhu08a, Zhu08b} and galaxy structures,
formations, and evolutions
\citep{Zhu97,MS98,Jin00,Kee01,KW01,CM03,Ofek03,RK05,Cha05,Cha06,Koo06,Tre06}.
Now several hundreds of strong lens systems produced by massive
galaxies have been discovered, but only $\sim 90$ galactic-scale
strong lenses with known redshift of source and image separation can
form well-defined samples useful for statistical analyses.

These well-defined strong lenses are particularly useful not only
for constraining the statistical properties of galaxies such as
optical region velocity dispersions \citep{Cha05,Cha06} and galaxy
evolutions \citep{CM03,Ofek03}, but also for constraining
cosmological parameters such as the present matter density
$\Omega_m$ and the equation of state of dark energy $w$
\citep{Cha03, Mit05}. For example, the CLASS statistical sample
\citep{Bro03,Cha03} containing 13 lenses which strictly satisfy
well-defined selection criteria was first extensively used by
\citet{Cha02} and \citet{Cha03}, who found $\Omega_m \approx 0.3$ in
a flat cosmology with non-evolving galaxy populations. Then this
CLASS sample was reanalyzed with the velocity dispersion function
(VDF) of early-type galaxies derived from the SDSS Data Release 1
(DR1 \citep{She03}) galaxies \citet{Mit05}.

In our paper, we summarize a complete lens redshift sample from
various imaging surveys including the Cosmic Lens ALL-Sky Survey
(CLASS; \citep{Bro03,Mye03}),the JVAS, the Sloan Digital Sky Survey
(SDSS; \citep{Ogu06}), the PMN-NVSS Extragalactic Lens Survey
(PANELS; \citep{Win01})and the Snapshot survey, which accumulates 29
galactic-scale lenses so far to form a well-defined radio-selected
lens sample. Newly measured J1620+1203 \citep{Kayo09} from SDSS is
also included.

One emphasis we need to make is the measurement of the VDF of
galaxies. \citet{Cha05} found that the \citet{She03} VDF of
early-type galaxies underestimated their abundance based on the
Wilkinson Microwave Anisotropy Probe (WMAP) 1st year cosmology
\citep{Spe03}. Then, with a new method of classifying galaxies
\citep{PC05}, \citet{Cho06} made a new VDF measurement of early-type
galaxies based on the much larger SDSS Data Release 5 (DR5;
\citet{Ade08}). Recently, \citet{Cha06b} have determined the VDF of
the late-type population using the Tully-Fisher relation and SIE
galaxy model, which matches relatively well that of \citet{She03}.
More recently, \citet{Cha10} introduced a correction term for high
velocity dispersions and used the Monte Carlo method to separately
generate the early-type and late-type galaxies and to derive a total
VDF for the entire population of galaxies (see also in
\citet{Bernardi10}). However, the simulated data points for the
total population of galaxies are not fitted well by the VDF of the
morphologically-typed galaxy populations, which might be due to the
errors in the adopted correlations between luminosity and velocity
\citep{Cha10}.

Moreover, strong lensing has also been extensively applied to
constrain dark energy, one of the most important issues of the
modern cosmology ever since the observations of type Ia supernovae
(SNe Ia) first indicating an accelerating expansion of the universe
at the present stage \citep{Riess98}. Among diverse dark energy
models, the most simple candidate for the uniformly distributed dark
energy is considered to be in the form of vacuum energy density or
cosmological constant ($\Lambda$). However, the cosmological
constant is always entangled with (i) fine tuning problem (present
amount of the dark energy is so small compared with fundamental
scale) and (ii) coincidence problem (dark energy density is
comparable to critical density today). Alternatively there exist
other choices, for example, an X-matter component, which is
characterized by an equation of state $p= w\rho $, where $-1\leq
w\leq 0$ \citep{Zhu98,Peebles03}. The goal of this work is to use
the lens redshift test combined with the revised VDF of all-type
galaxies based on the SDSS DR5 data to constrain two cosmological
models.

This paper is organized as follows. In Section~\ref{sec:method} we
briefly describe the analysis method including assumptions about the
lens population. We then present the lens redshift data from various
surveys in Section~\ref{sec:data}. We further introduce two
prevalent cosmologies and show the results of constraining
cosmological parameters using MCMC method in
Section~\ref{sec:result}. Finally, we conclude and make a discussion
in Section~\ref{sec:conclusion}.

\section{Lens redshift test}
\label{sec:method}

The statistical lensing model used in this paper incorporates the
(differential) lensing probabilities of specific image
multiplicities for the multiply-imaged sources using the SIE lens
model. The primary assumption we make is that the distribution of
early-type galaxies in luminosity is given by the Schechter (1976)
form
\begin{equation}
\tilde{\phi}(L) dL=\tilde{\phi}_* \left(\frac{L}{L_*}\right)^{\tilde{\alpha}}
      \exp\left(-\frac{L}{L_*}\right) \frac{dL}{L_*}.
\end{equation}
Considering a power-law relation between luminosity ($L$) and
velocity dispersion ($\sigma$), i.e. $\frac{L}{L_*} =
\left(\frac{\sigma}{\sigma_*}\right)^{\gamma}$, we can describe the
distribution of galaxies in velocity dispersion in the form of the
modified Schechter function $\phi(\sigma)$ \citep{She03, Mit05}
\begin{equation} dn = \phi(\sigma) d\sigma = \phi_*
  \left(\frac{\sigma}{\sigma_*}\right)^{\alpha}
   \exp\left[-\left(\frac{\sigma}{\sigma_*}\right)^{\beta}\right]
   \frac{\beta}{\Gamma(\alpha/\beta)} \frac{d\sigma}{\sigma},
\label{VDF} \end{equation} where $\phi_*$ is the integrated number
density of galaxies, $\sigma_*$ is the characteristic velocity
dispersion, with the following relations: $\alpha = (\tilde{\alpha}
+ 1) \gamma$, $\beta = \gamma$, and $\phi_* = \tilde{\phi}_*
\Gamma(\tilde{\alpha} + 1)$.

Following Eq.~(\ref{VDF}), the particular differential probability
that a source at redshift $z_s$ be multiply imaged by a distribution
of galaxies at redshift $z_l$ with a image separation $\Delta\theta$
can be written as:
\begin{eqnarray}
\delta p & \equiv & \frac{d^2p}{dzd(\Delta\theta)}/\frac{dp}{dz} \nonumber \\
  & = & \frac{1}{2} \frac{\beta}{\Gamma[(\alpha+4)/\beta]}
     \frac{1}{\Delta\theta_*}
  \left(\frac{\Delta\theta}{\Delta\theta_*}\right)^{\alpha/2+1} \exp\left[-\left(\frac{\Delta\theta}{\Delta\theta_*}\right)^{\beta/2}\right],
  \label{deltap}
\end{eqnarray}
where $\Delta\theta_*$ is  the characteristic image separation given
by
\begin{equation}
\Delta\theta_* = \lambda 8 \pi \frac{D(z,z_s)}{D(0,z_s)}
                 \left(\frac{\sigma_*}{c}\right)^2
\end{equation}
and $D(z_1,z_2)$ is the angular-diameter distance between redshifts
$z_1$ and $z_2$. On the hypothesis that galaxies are not biased
toward oblate or prolate shape, we choose the dynamical
normalization factor $\lambda \approx 1$ for the singular isothermal
ellipsoid (SIE) model \citep{Cha03,Cha05}. Though isothermal mass
model would, in general, be too simplified to do accurate modelling
of individual lenses \citep{Cha02}, it is accurate enough as
first-order approximations to the mean properties of galaxies for
the analyses of statistical lensing
(e.g.\citep{Koc93,Mao94,Rix94,Koc96,Kin97,Fassnacht98,RK05}).

Following \citet{Cha03}, the likelihood $\mathcal{L}$ of the
observed image separations for $N_{\rm{L}}$ multiply-imaged sources
reads
\begin{equation}
\ln \mathcal{L} = \sum_{l=1}^{N_{\rm{L}}} \omega_l^{(g)}  \ln \delta p_l.
\label{lnL}
\end{equation}
Here $\delta p_l$ is the particular differential probability for the
$l$-th multiply-imaged source. $\omega_l^{(g)}$ ($g = e$, $s$) is
the weight factor given to the early-type or the late-type
populations, which satisfies $\omega_l^{(e)} + \omega_l^{(s)} = 1$.
If the lensing galaxy type is unknown, we use $\omega_l^{(g)} =
\delta p_{l}^{(g)}/ [\delta p_{l}^{(e)}+ \delta p_{l}^{(s)}]$ ($g =
e$, $s$) \citep{Cha03}. Accordingly, a ``$\chi^2$''is defined as:
\begin{equation}
\chi^2 = -2 \ln \mathcal{L}.
\label{chi}
\end{equation}

Notice that the $\chi^2$ here is free of the dimensionless Hubble
constant $h$, which makes it possible to perform as an individual
cosmological probe besides SNeIa, CMB, BAO etc. The `best-fit' model
parameters are determined by minimizing the $\chi^2$
[Eq.(\ref{chi})] and confidence limits on the model parameters are
obtained by using the usual $\Delta\chi^2$ ($\equiv \chi^2 -
\chi^2_{\rm min}$) static, where $\chi^2_{\rm min}$ is the global
minimum $\chi^2$ for the best-fit parameters. Our calculating method
is based on the publicly available package CosmoMC \citep{Lewis02}.

On the side of the measurement of VDF, the first direct measurement
of the VDF of early-type galaxies was the SDSS DR1 \citep{She03}:
\begin{eqnarray}
 (\phi_*,\sigma_*,\alpha,\beta)_{\rm DR1}
  & = & [(4.1 \pm 0.3) \times 10^{-3} h^3 {\rm Mpc}^{-3}, \nonumber  \\
  &   & 88.8 \pm 17.7 \rm \,km~s^{-1}, \nonumber \\
  &   & 6.5 \pm 1.0, 1.93 \pm 0.22],
\label{DR1}
\end{eqnarray}
Then \citet{Cho06} obtained a new VDF based on the much larger SDSS
DR5:
\begin{eqnarray}
 (\phi_*,\hspace{0.2cm} \sigma_*,\alpha, \beta)_{\rm DR5}
  & = & [8.0 \times 10^{-3}h^3 {\rm Mpc}^{-3}, \nonumber  \\
  &   & 161 \pm 5 \rm \,km~s^{-1}, \nonumber \\
  &   & 2.32 \pm 0.10, 2.67 \pm 0.07].
\label{DR5}
\end{eqnarray}
Obviously, the revised DR5 VDF, which has been proved to provide an
efficient way to constrain dark energy models combined with
gravitational lensing statistics \citep{Zhu08a}, is quite different
from the DR1 VDF in the characteristic velocity dispersion at
1$\sigma$. While early-type galaxies dominate strong lensing,
late-type galaxies cannot be neglected. However, for the late-type
galaxy population, the direct measurement of the VDF is complicated
by the significant rotations of the disks. \citet{Cha06b} estimated
all the parameters of equation~(\ref{VDF}) for the late-type
population using the Tully-Fisher relation and SIE galaxy model:
\begin{eqnarray}
 (\phi_*,\sigma_*,\alpha,\beta)_{\rm late}
  & = & [1.13 \times 10^{-1} h^3 {\rm Mpc}^{-3}, \nonumber  \\
  &   & 133 \rm \,km~s^{-1}, 0.3, 2.91],
\label{late}
\end{eqnarray}
which matches relatively well that of \citet{She03} who determined
$\sigma_*^{\rm(late)}$ using a Tully-Fisher relation.

Currently, it is found that simple evolutions do not significantly
affect the lensing statistics if all galaxies are early-type
\citep{Mao94, Rix94, Mit05,CN07}. Many previous studies on lensing
statistics without evolutions of the velocity function have also got
appealing results that agree with the galaxy number counts
\citep{Im02} and the redshift distribution of lens galaxies
\citep{Cha03, Ofek03}. However, in this paper, we consider galaxy
evolutions both for early-type and late-type galaxies with a recent
semi-analytical model of galaxy formation \citep{Kan05,Cha06}:
\begin{equation}
 \phi_*(z)=\phi_{*,0} (1+z)^{\nu_n};\hspace{0.2cm}
 \sigma_*(z)=\sigma_{*,0} (1+z)^{\nu_v}.
\label{evmod}
\end{equation}
with the best-fit parameters of \citet{Kan05}:
$(\nu_n,\hspace{0.2cm} \nu_v) = (-0.229, -0.01)$ for early-type and
$(\nu_n,\hspace{0.2cm} \nu_v) = (1.24, -0.186)$ for late-type
galaxies.

\section{Lens redshift data} \label{sec:data}
Large systematic surveys of gravitationally lensed quasars provide a
large statistical lens sample appropriate for studying cosmology. In
this section, we summarize a sample both from CLASS, SDSS
observations and recent large-scale observations of galaxies, which
will be used as the input for the statistical lensing model
described in section \ref{sec:method}. Two main sources of the lens
redshift data are the Cosmic Lens All-Sky Survey and the Sloan
Digital Sky Survey Quasar Lens Search.\\

\textbf{A. CLASS}\\

As the largest completed galactic mass scale gravitational lens
search project, the Cosmic Lens All-Sky Survey, along with its
predecessor project: the Jodrell-Bank--VLA Astrometric Survey (JVAS)
has confirmed 22 multiply-imaged systems out of a total of 16521
radio sources \citep{Bro03}. Out of the entire CLASS sample
including the JVAS sources, a subsample of 8958 sources containing
13 multiply-imaged systems constitutes a statistical sample that
satisfies well-defined observational selection criteria
\citep{Bro03}, which can be used for the statistical analysis of
gravitational lensing in this paper.

In this work we need galactic-scale strong lens samples that satisfy
well-defined observational selection criteria. With the well-defined
selection criteria from \citet{Bro03,Cha03}, two out of the four
multiply-imaged sources in the JVAS, 0414+054 with a too steep
spectral index and 1030+074 with the fainter-to-brighter image
flux-density ratio of two images less than 0.1, are excluded from
the final CLASS statistical sample. Meanwhile, we stress that the
measured lens redshifts, source redshifts, image separations, image
multiplicities and the lensing galaxy types (if determined), as
shown in table~\ref{tab:data}, are all needed through the likelihood
function defined in Section~\ref{sec:method} to constrain
cosmological parameters. Under this criterion, for the CLASS
sources, B0850+054, B0445+123, B0631+519, B1938+666 are clearly
excluded due to their unknown source redshifts, while for B0739+366,
J2004.1349, redshifts both for source and lens are unavailable
\footnote{We discard lens systems that do not have measured lens or
source redshifts, which may possibly cause biases. For many
multiply-imaged sources without measured source redshifts, a
possible strategy is to take $z_s = 2$, which is the mean source
redshift for the multiply-imaged sources with measured source
redshifts. However, in order to ensure the accuracy of constraint,
we choose to abandon such a choice in our paper.}. Moreover, the
measured image separations of 1359+154, 1608+656 and 2114+022 are
not used because the observed angular sizes are due to multiple
galaxies within their critical radii.\\

\textbf{B. SDSS}\\

The Sloan Digital Sky Survey Quasar Lens Search (SQLS;
\citet{Ogu06}) is a photometric and spectroscopic survey covering
nearly a quarter of the entire sky \citep{Yor00}. We try to find
suitable lens samples from the optical quasar catalog of the Sloan
Digital Sky Survey (SDSS; \citet{Yor00}). The first complete lens
sample from Data Release 3 selected from 22,683 low-redshift
($0.6<z<2.2$) is provided in \citet{Inada08}. It consists of 11
lensed quasars satisfying the following well-defined selection
criteria: 1)The image separation is  between $1''$ and $20''$ with
quasars brighter than $i=19.1$; 2) The flux ratio of faint to bright
images is greater than $10^{-0.5}$ for double lenses. In this paper,
an additional cut is applied to select a appropriate subsample: the
lensing galaxy should be fainter than the quasar components, because
a too bright lens galaxy will strongly affect the colors of the
quasars \citep{Richards02}. Four lensed quasars, Q0957+561, SDSS
J1004+4112, SDSS J1332+0347 and SDSS J1524+4409 are removed with
this cut. Like the CLASS, due to the absence of lens redshifts, two
more quasars, SDSS J1001+5027 and SDSS J1021+4913 are excluded.
However, we successfully add four lens redshift data including SDSS
J1620+1203, one of the eight newly discovered and confirmed
two-image lensed quasars by SDSS Quasar Lens Search \citep{Kayo09}.

In table~\ref{tab:data}, we summarize 29 Strongly-lensed Sources
(redshifts both for sources and lenses, as well as the largest image
separations and galaxy types) from the CLASS (JVAS), the SDSS, the
PANELS and the Snapshot, which formulate a well-defined combined
statistical sample for model constraints.

\begin{table*}
\caption{\label{tab:data}Summary of Strongly-lensed Sources. The
properties of the strongly lensed systems from the Snapshot optical
survey and the CLASS (\citealt{Bro03}) and PANELS radio surveys are
revised from \citet{Koc96} and \citet{Cha03, Cha05}. References are
the following: 1 - the CASTLES website
({\mbox{http://cfa-www.harvard.edu/castles/}});  2 -
\citet{Cha03,Cha05}; 3 - \citet{Koc96};  4 - \citet{Kin97}; 5 -
\citet{Inada05} 6 - \citet{Eigenbrod07}; 7 - \citet{Oscoz97}; 8 -
\citet{Lubin00}; 9 - \citet{Inada03a}; 10 - \citet{Ofek03}; 11 -
\citet{Eigenbrod06a}; 12 - \citet{Inada03b}; 13 - \citet{Eigenbrod
06b}; 14 - \citet{Oguri04}; 15 - \citet{Walsh79}; 16 -
\citet{Young81}; 17 - \citet{Morokuma07}; 18 - \citet{Oguri08}; 19 -
\citet{Fassnacht98}; 20 - \citet{McKean07}; 21 - \citet{Surpi03}; 22
- \citet{Kayo09}; 23 - \citet{Win01} \label{complete}}
\begin{tabular}{clllllll}

\hline
  &    & Source & Lens & Maximum & Number &Lensing \\
 Source & Survey  & redshift   & redshift
 & image  & of & galaxy(-ies) \\
 &        & ($z_{s}$)  & ($z_{l}$)
 & separation ($''$)  & images &  type \\
\hline
 B0414+054  & CLASS & 2.64   & 0.96  & 2.03  & 4  & early-type (1, 2)  \\
 B0712+472$^\dagger$  & CLASS &  1.34  & 0.41  & 1.27  & 4 & early-type (1, 2)\\
 B1030+074  & JVAS  &  1.535 & 0.599 & 1.56   & 2  & 2Gs (E+?) (1, 2)\\
 B1422+231$^\dagger$  & JVAS  &  3.62  & 0.34  & 1.28   & 4 & early-type (1, 2)\\
J1632$-$0033 & PANELS & 3.42  &  1  & 1.47   & 2  & early-type (1, 2) \\
 J1838$-$3427 & PANELS & 2.78  & 0.36 & 1.0   & 2  & early-type (1, 2) \\
B1933+503$^\dagger$  & CLASS &  2.62  & 0.755 & 1.17   & 4 & early-type (1, 2) \\
Q0142-100 & Snapshot & 2.72 & 0.49 & 2.23 & 2 & early-type (1, 3) \\
PG1115+080 & Snapshot & 1.72 & 0.31& 2.43& 4 & early-type (1, 3)\\
B1938+666 & CLASS   &1.8 & 0.88 & 0.91 & 4+2+R & early-type (1, 2, 4)\\
J0246$-$0825 & SDSS & 1.685 & 0.723 &1.04 &2 & early-type (5, 6)\\
SBS0909+523 &SDSS  &1.377 &0.83 &1.11 &2 & early-type  (7, 8)\\
 J0924+0219 &SDSS  &1.523 &0.393  &1.78 &4 & early-type (9, 10, 11)\\
 J1226$-$0006 &SDSS  &1.125 &0.517 &1.24 & 2  & early-type  (12, 13)\\
 J1335+0118 &SDSS  &1.571 &0.440 &1.57 &2  &early-type (14, 13)\\
 Q0957+561  &SDSS  &1.413 &0.36 &6.17 &2 &early-type (15, 16)\\
 J1332+0347 &SDSS  &1.438 &0.191  &1.14 &2 &early-type  (17)\\
 J1524+4409 &SDSS  &1.210 &0.310  &1.67 &2 &early-type (18)\\
 0712+472   &CLASS & 1.34 &0.41   &1.27 &4 &early-type (19)\\
 B1359+154  & CLASS & 3.24  & 1  & 1.67  & 6    & 3Gs (E+?+?) (1, 2) \\
 B2045+265  &CLASS &4.3  &0.87   &1.91 &4 & 2Gs (E+?) (2, 20)\\
 B1608+656  &CLASS &1.39 &0.64   &2.08 &4 & 2Gs (E+L) (2, 21)\\

\hline
B0128+437 & CLASS & 3.12 & 1.15 & 0.55 & 4 &unknown (6, 14) \\
B1152+199$^\dagger$  & CLASS &  1.019 & 0.439 & 1.56   & 2  & 2Gs [?(E)a+?] (1, 2)\\
Q1208+1011 & Snapshot & 3.80 & 1.13 & 0.48 & 2  &unknown (5, 7, 8)\\
J1620+1203 &SDSS  &1.158  &0.398  &2.765 &2 &unknown  (22)\\

\hline
B0218+357 & CLASS & 0.96 & 0.68 & 0.33 & 2 &late-type (1, 2) \\
B1600+434 & CLASS & 1.59 & 0.41 & 1.38 & 2 &late-type (1, 2) \\
J0134.0931 & PANELS &  2.23 & 0.76 & 0.68  & 5+2  &2Gs (L?+L?) (1, 2, 23)\\

\hline
\end{tabular}

\end{table*}

\section{DE models and constraint results}
\label{sec:result}

In what follows, we choose two popular dark energy models and
examine whether they are consistent with the lens redshift data
listed above.
\begin{enumerate}
\item Cosmological constant model.
\item Dark energy with constant equation of state.
\end{enumerate}

Both of the two classes of models are currently viable candidates to
explain the observed cosmic acceleration. Unless stated otherwise,
throughout the paper we calculate the best fit values found, and
vary the parameters within their 2$\sigma$ uncertainties for either
class of model. Next, we shall outline the basic equations
describing the evolution of the cosmic expansion in both dark energy
models and calculate the best-fit parameters.

\subsection{Constraint on the standard cosmological model ($\Lambda$CDM)}
\subsubsection {Constraint from the lens redshift data}
In the simplest scenario, the dark energy is simply a cosmological
constant, $\Lambda$, i.e.\ a component with constant equation of
state $w=p/\rho=-1$. If flatness of the FRW metric is assumed, the
Hubble parameter according to the Friedmann equation is:
\begin{equation}
\left(\frac{H}{H_0}\right)^2=\frac{\Omega_m}{a^3} +  \Omega_\Lambda,
\end{equation}
where $\Omega_m$ and $\Omega_\Lambda$ parameterize the density of
matter and cosmological constant, respectively. Moreover, in the
zero-curvature case ($\Omega=\Omega_m+\Omega_\Lambda=1$), this model
has only one independent parameter: ${\theta}=\Omega_{\Lambda}$.

\begin{figure}
\begin{center}
\includegraphics[width=0.4\hsize]{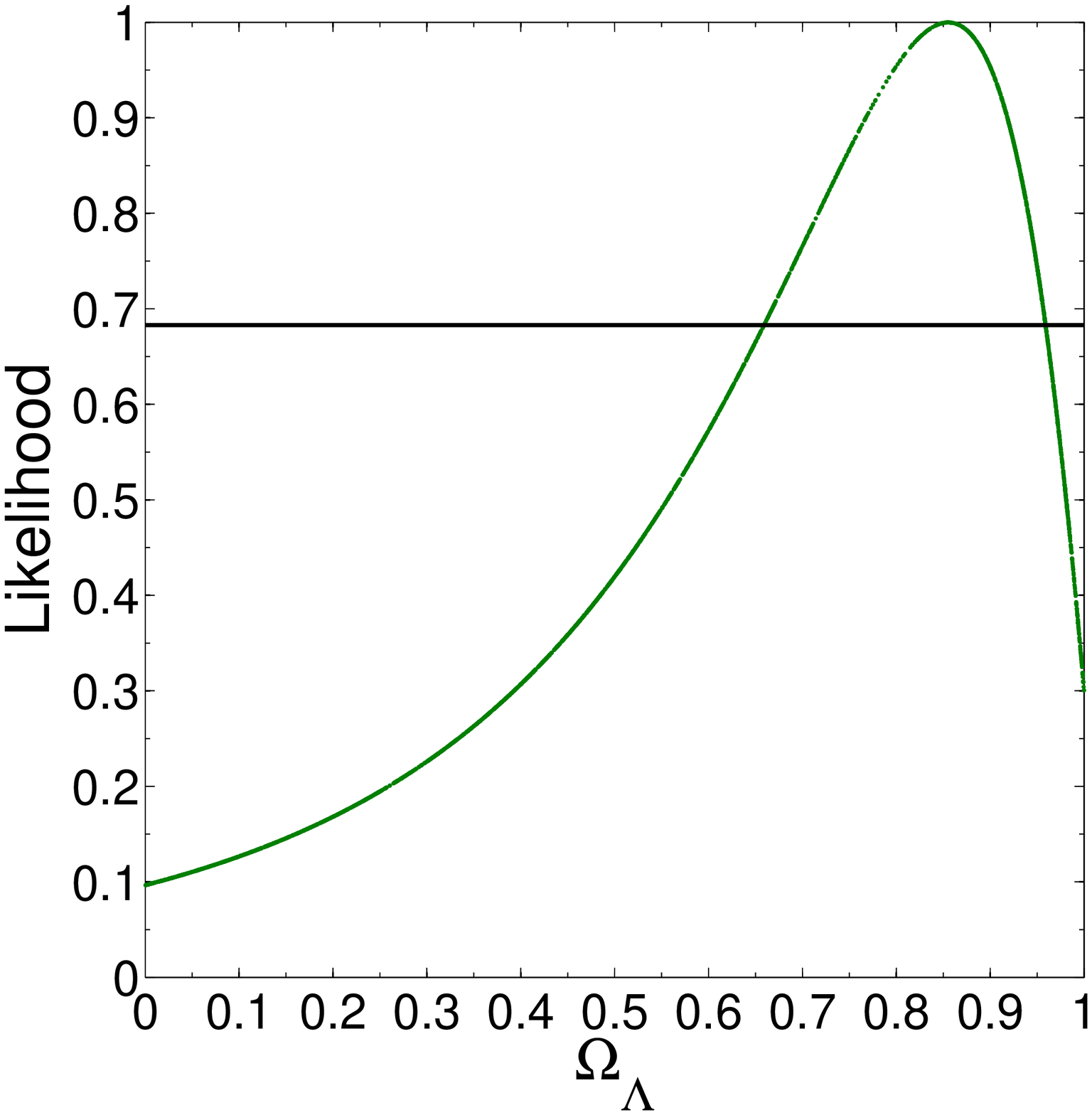}
\end{center}
\caption{\textbf{The likelihood distribution function for the $\Lambda$ CDM model constrained by the lens redshift data.}
\label{1}}
\end{figure}

We plot the likelihood distribution function for this model in
Fig.~\ref{1}. The best-fit value of the parameter is:
$\Omega_\Lambda=0.85^{+0.11}_{-0.18}$. It is obvious that the lens
redshift data only give a relatively weak constraint on the model
parameter $\Omega_\Lambda$, though the universally recognized value
of $\Omega_\Lambda=0.75$ is still included at 68.3\% CL (1$\sigma$).
To make a comparison, it is necessary to refer to the previous
results: the current best fit value from cosmological observations
is: $\Omega_\Lambda=0.73\pm 0.04$ in the flat case \citep{Davis07},
which is in relatively stringent accordance with our result.
Moreover, \citet{Komatsu09} gave the best-fit parameter:
$\Omega_{m}=0.274$ for the flat $\Lambda$ CDM model from the WMAP
5-year results with the BAO and SN Union data. We find that the
constraint result from the lens redshift data is marginally
consistent with the previous works above.

\subsubsection {Joint analysis with BAO and CMB}
For the BAO data, the parameter $\mathcal{A}$ is used, which, for a
flat universe can be expressed as
\begin{equation}
\mathcal{A}=\frac{\sqrt{\Omega_m}}{(H(z_{BAO})/H_0)^{1/3}}\bigg[\frac{1}{z_{BAO}}\int_0^{z_{BAO}}\frac{dz}{H(z)/H_0}\bigg]^{2/3}\;,
\end{equation}
where $z_{BAO}=0.35$ and $\mathcal{A}$ is
$\mathcal{A}=0.469\pm0.017$ from the SDSS \citep{Eisenstein05}.

For the CMB data, the shift parameter $\mathcal{R}$ is used, which
may provide an effective way to constrain the parameters of dark
energy models due to its large redshift distribution. Derived from
the CMB data, $\mathcal{R}$ takes the form as
\begin{eqnarray}
\mathcal{R} &=&\sqrt{\Omega_{m}}\int_{0}^{z_{\mathrm{CMB}}}\frac{dz}{%
H(z)/H_{0}}  \label{shift parameter},
\end{eqnarray}%
where $z_{CMB}=1090$ \citep{Komatsu09} is the redshift of
recombination and the 5-year WMAP data give $\mathcal{R} = 1.710 \pm
0.019$ \citep{Komatsu09}.

\begin{figure}
\begin{center}
\includegraphics[width=0.4\hsize]{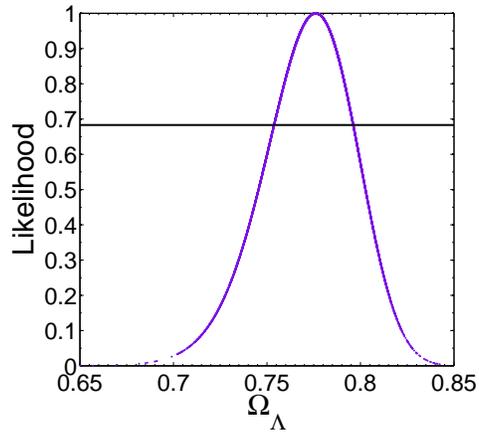}
\end{center}
\caption{The likelihood distribution function for the $\Lambda$ CDM model constrained by the lens redshift data combined with CMB.
\label{5}}
\end{figure}

\begin{figure}
\begin{center}
\includegraphics[width=0.4\hsize]{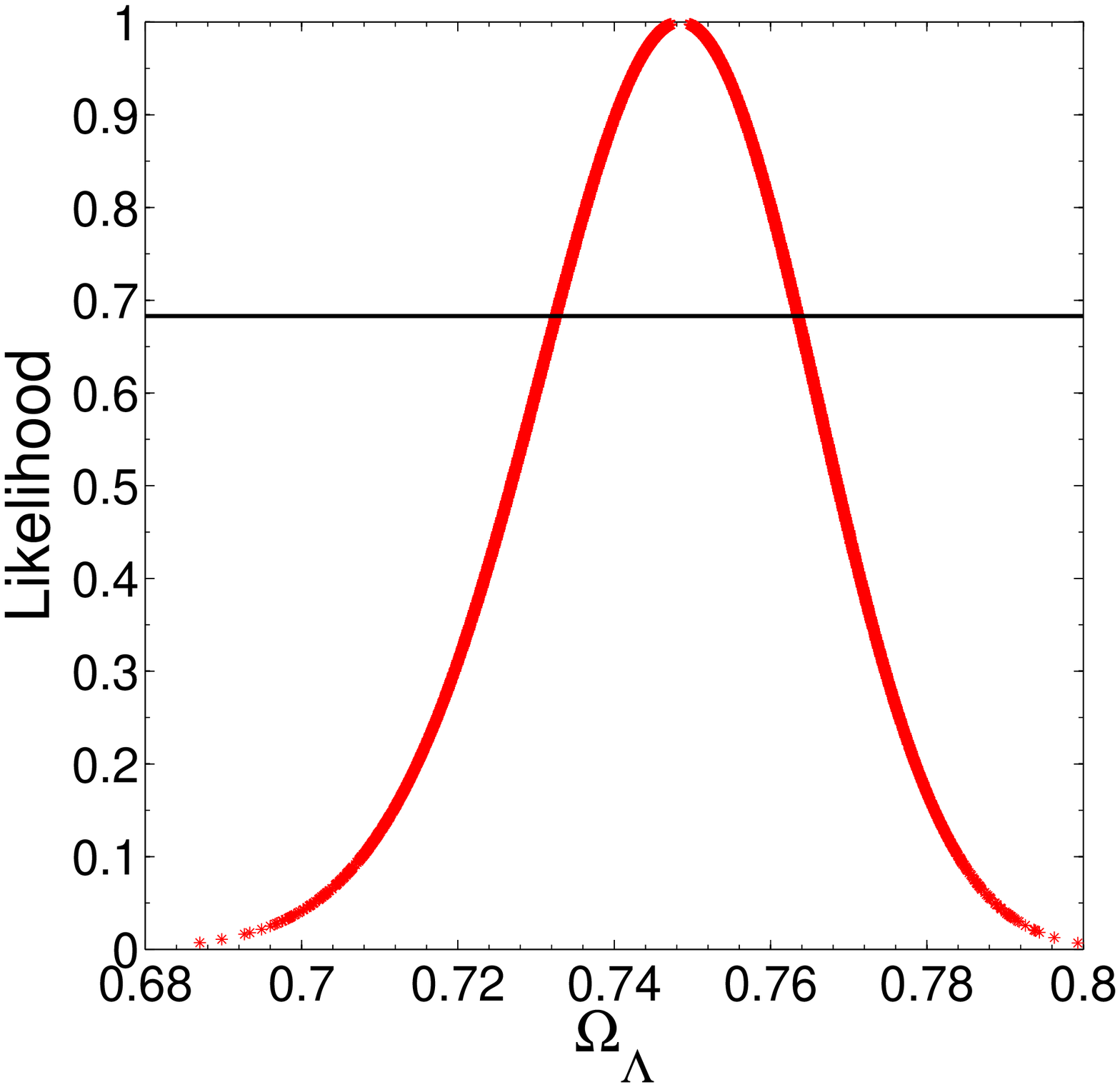}
\end{center}
\caption{The likelihood distribution function for the $\Lambda$ CDM model constrained by the lens redshift data combined with CMB and BAO.
\label{2}}
\end{figure}

In Fig.~\ref{5} and \ref{2}, we show the result by combining the
lens redshift data with CMB and CMB+BAO, respectively. The best-fit
parameter is: $\Omega_\Lambda=0.78^{+0.02}_{-0.03}$ and
$\Omega_\Lambda=0.75^{+0.02}_{-0.02}$, a relatively satisfactory
result consistent with that of \citet{Li10}. Compared with
Fig.~\ref{1}, it is apparent that $\Omega_\Lambda$ is more tightly
constrained with the joint data sets.

\subsection{Dark energy with constant equation of state ($w$)}
\subsubsection {Constraint from the lens redshift data}
If allowing for a deviation from the simple $w=-1$, a component with
a constant value for the equation of state could be introduced. The
accelerated expansion can be obtained when $w<-1/3$. In a
zero-curvature universe, the Hubble parameter for this generic dark
energy component with density $\Omega_x$ then becomes:
\begin{equation}
\left(\frac{H}{H_0}\right)^2=  \frac{\Omega_m}{a^3} + \frac{\Omega_x}{a^{3(1+w)}} .
\end{equation}
Obviously, when flatness is assumed
($\Omega=\Omega_m+\Omega_\Lambda=1$), it is a two-parameter model
with the model parameters: ${\theta}=\{\Omega_x,~w\}$.

\begin{figure}
\begin{center}
\includegraphics[width=0.4\hsize]{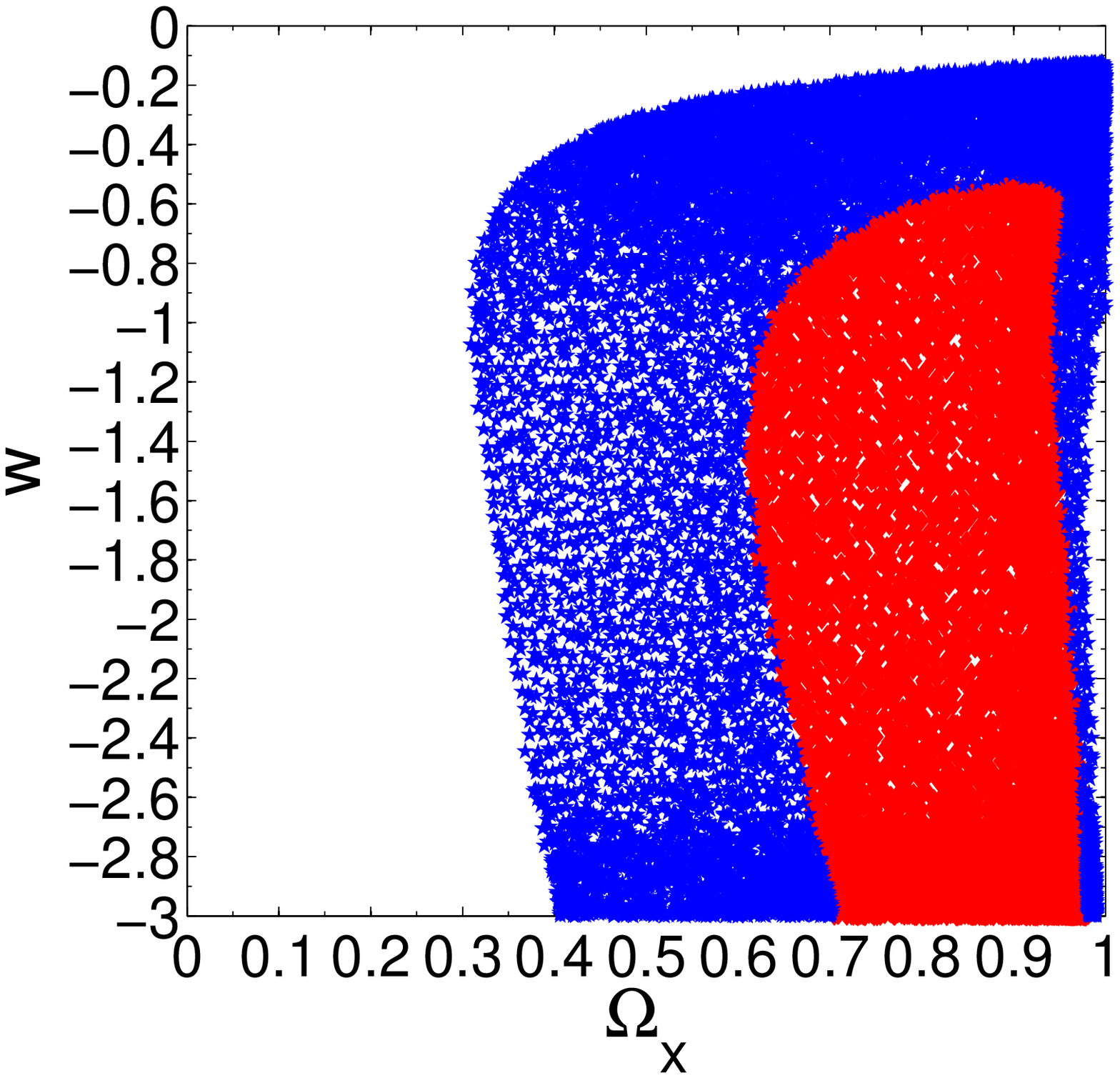}
\end{center}
\caption{The $w$ model constrained by the lens redshift data: likelihood contours at
68.3\% and 95.4\% CL in the $\Omega_x-w$ plane.
\label{3}}
\end{figure}

For the redshift data only, the best-fit values of the parameters
are: $\Omega_x=0.78, w=-1.15$. Fig.~\ref{3} shows the confidence
limits in the $\Omega_x-w$ plane. On the one hand, we have $w<-0.52$
at 68.3\% CL, which is quite different from the result of
\citet{Cha06b}; on the other hand, the Einstein's cosmological
constant ($w=-1$) is still favored within 1$\sigma$ error region.
Therefore, it seems that the present lens redshift data do not
necessarily favor the phantom DE model with $w<-1$ \citep{Cald}.
However, it is still interesting to see whether this remains so with
future lager and better lens redshift data.

\subsubsection {Joint analysis with BAO and CMB}
\begin{figure}
\begin{center}
\includegraphics[width=0.4\hsize]{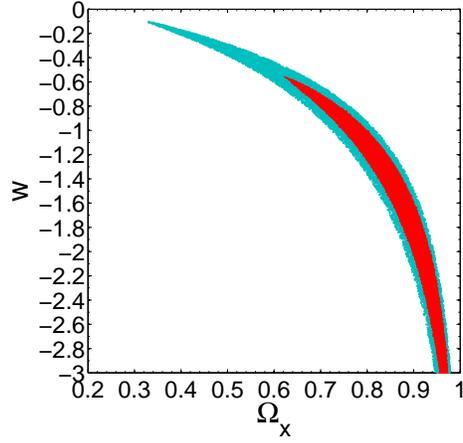}
\end{center}
\caption{The $w$ model constrained by the lens redshift data combined with CMB: likelihood contours at
68.3\% and 95.4\% CL in the $\Omega_x-w$ plane.
\label{6}}
\end{figure}

\begin{figure}
\begin{center}
\includegraphics[width=0.4\hsize]{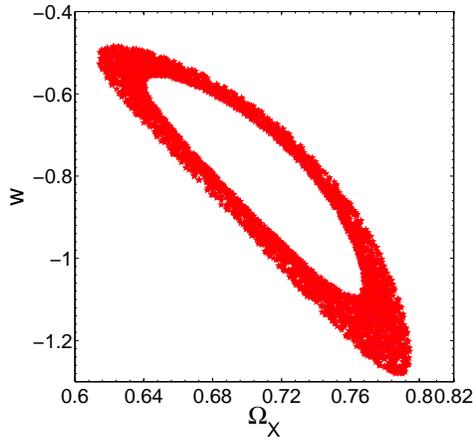}
\end{center}
\caption{The $w$ model constrained by the lens redshift data combined with CMB and BAO: likelihood contours at
68.3\% and 95.4\% CL in the $\Omega_x-w$ plane.
\label{4}}
\end{figure}

In Fig.~\ref{6} and \ref{4}, we plot the likelihood contours with
the joint data by combining the lens redshift data with CMB and
CMB+BAO in the $\Omega_x-w$ plane. The best-fit parameters are: $
\Omega_x=0.80^{+0.17}_{-0.17}, w=-1.12^{+0.57}_{-1.88}$ and $
\Omega_x=0.71^{+0.07}_{-0.07}, w=-0.78^{+0.22}_{-0.34}$. Notice that
both $\Omega_x$ and $w$ are more stringently constrained with the
joint observational data. Meanwhile, the currently preferred values
of $w$ in this model still include the cosmological constant case:
$w=-1.01\pm 0.15$ \citep{Davis07}. Therefore, when the equation of
state does not depend on the redshift, the dark energy is consistent
with a flat cosmological constant model within 1$\sigma$ error
region.

\section{Conclusion and discussion }
\label{sec:conclusion} Recent new observations, such as SNeIa,
Wilkinson Microwave Anisotropy Probe (WMAP) \citep{Komatsu09} and
Baryon Acoustic Oscillations (BAO) \citep{Percival07}, the time
drift of subtended angles \citep{Zhang09}, the updated Gamma-ray
bursts (GRB) \citep[e.g.,][]{Gao10,Liang10} have provided many
robust tools to study the dynamical behavior of the universe.
However, it is still important to use other different probes to set
bounds on cosmological parameters. In this work, we have followed
this direction and used the distribution of gravitationally-lensed
image separations observed in the Cosmic Lens All-Sky Survey
(CLASS), the PMN-NVSS Extragalactic Lens Survey (PANELS), the Sloan
Digital Sky Survey (SDSS) and other surveys to constrain
cosmological models with the new measurements of the velocity
dispersion function of galaxies based on the SDSS DR5 data and
recent semi-analytical modeling of galaxy formation. Two dark energy
models ($\Lambda$CDM and constant $w$) are considered under a flat
universe assumption.

For the zero-curvature $\Lambda$ CDM model, although the lens
redshift data can not tightly constrain the model parameter
$\Omega_\Lambda=0.85^{+0.11}_{-0.18}$, a stringent constraint can be
obtained by combining the lens redshift data with the comic
macrowave background data: $\Omega_\Lambda=0.78^{+0.02}_{-0.03}$ and
the baryonic acoustic oscillation peak data as well:
$\Omega_\Lambda=0.75^{+0.02}_{-0.02}$. Furthermore, we consider a
flat cosmology with a constant $w$ dark energy. For the lens
redshift data, we have $w<-0.52$ at 68.3\% CL, a result different
from that of \citet{Cha06b} with $w<-1.2$ (68.3\% CL), therefore,
this strong lensing data do not necessarily favor a super-negative
equation of state for dark energy. However, the Einstein¡¯s
cosmological constant ($w=-1$) is still included within 1$\sigma$
error region. Likewise, adding CMB and CMB+BAO does lead to further
improvements in parameter constraints with
$\Omega_x=0.80^{+0.17}_{-0.17}, w=-1.12^{+0.57}_{-1.88}$ and $
\Omega_x=0.71^{+0.07}_{-0.07}, w=-0.78^{+0.22}_{-0.34}$,
respectively. Therefore, it indicates that the cosmological constant
model is still the best one to explain this lens redshift data, a
conclusion in accordance with the previous works \citep{Davis07} and
the results from the WMAP and the large-scale structures in the SDSS
luminous red galaxies \citep{Spe03,Tegmark04,Eisenstein05}.

However, we also notice that, firstly, the implementation of
singular isothermal ellipsoid model (SIE) may be a source of
systematic errors. For example, a lens ellipticity of 0.4 can lead
to a difference of $\Delta \Omega_m \approx -0.05$ compared with the
spherical case due to the variation of magnification bias and cross
section \citep{Hut05}. Secondly, as for the source of the lens
redshift data, we simply discard lens systems that do not have
measured lens or source redshifts in this paper. This could also
possibly bring biases and will be considered in our future work.
Thirdly, though the lens redshift test applied in this paper is free
from the magnification bias arising from the uncertain source
counts, it may also lose the statistical power of absolute lensing
probabilities (or "lensing rates"). Lastly, though we have used the
VDF of galaxies based on a much larger SDSS Data Release in the
strong lensing statistics, the accuracy of measurements on relevant
parameters of VDF may also make a difference. Hopefully, Large new
samples of strong lenses will be expected to be obtained in future
wide-field imaging surveys such as the Large Synoptic Survey
Telescope (LSST; \citet{Ivezic00}) etc and within a few decades next
generation observation tools such as the Square Kilometre Array
(e.g., \citet{Bla04}) will also improve the precision of lensing
statistics by orders of magnitude. Therefore, the lens redshift test
can play an important role in uncovering the physical processes of
galaxy formation and universe evolution with much larger and better
lens redshift data.

Summarizing, from the above discussion we may safely arrive at a
conclusion that the results from the observational lens redshift
data are relatively agreeable and furthermore the lens redshift test
can be seen as a future complementarity to other cosmological
probes.

\section*{Acknowledgments}
This work was supported by the National Natural Science Foundation
of China under the Distinguished Young Scholar Grant 10825313 and
Grant 11073005, the Ministry of Science and Technology national
basic science Program (Project 973) under Grant No.2007CB815401, the
Fundamental Research Funds for the Central Universities and
Scientific Research Foundation of Beijing Normal University.

\end{document}